\documentclass[11pt]{article}

\usepackage{graphicx,a4,amsmath,amsfonts,amsthm,amssymb,url}

\newcommand{\E}{\mathsf{I\!E}}
\newcommand{\R}{\mathbb{R}}

\renewcommand{\P}{\mathbb{P}}
\newcommand{\var}{\mathrm{var}}
\newcommand{\argmax}{\mathrm{argmax}}
\newcommand{\argmin}{\mathrm{argmin}}

\newcommand{\supp}{\mathrm{supp}}
\newcommand{\op}[1]{\mathbf{#1}}
\newcommand{\opgreek}[1]{\boldsymbol{#1}}
\renewcommand{\vec}[1]{\mathbf{#1}}
\newcommand{\abs}[1]{\lvert#1\rvert}
\newcommand{\lag}{\langle}
\newcommand{\rag}{\rangle}
\newcommand{\conj}{\overline}
\newcommand{\trace}{\mathrm{tr}}
\newcommand{\norm}[1]{\lVert#1\rVert}
\newtheorem{prop}{Proposition}
\newtheorem{lemma}{Lemma}
\newtheorem{defin}{Definition}
\newcommand{\whiten}{\check}
\newcommand{\overwh}{\breve}

\title{A learning approach to the detection of gravitational wave transients} 
\author{E. Chassande-Mottin\\
Equipe ILGA-Virgo (CNRS-EP 2122), Observatoire de la C\^ote d'Azur\\ 
BP 4229 F-06304 Nice Cedex 4 France} 
\date{}

\begin{document}

\maketitle

\abstract{We investigate the class of quadratic detectors (i.e., the statistic is a bilinear
  function of the data) for the detection of poorly modeled gravitational transients of short
  duration. We point out that all such detection methods are equivalent to passing the signal
  through a filter bank and linearly combine the output energy. Existing methods for the choice of
  the filter bank and of the weight parameters (to be multiplied to the output energy of each filter
  before summation) rely essentially on the two following ideas~: (\textit{i}) the use of the
  likelihood function based on a (possibly non-informative) statistical model of the signal and the
  noise \cite{anderson01:_excess,vicere02:_optim}, (\textit{ii}) the use of Monte-Carlo simulations
  for the tuning of parametric filters to get the best detection probability keeping fixed the false
  alarm rate \cite{pradier01:_effic,arnaud99:_detec}.
  
  We propose a third approach according to which the filter bank is ``learned'' from a set of
  training data. By-products of this viewpoint are that, contrarily to previous methods,
  (\textit{i}) there is no requirement of an explicit description of the probability density
  function of the data when the signal is present and (\textit{ii}) the filters we use are
  non-parametric. The learning procedure may be described as a two step process ~: first, estimate
  the mean and covariance of the signal with the training data; second, find the filters which
  maximize a contrast criterion referred to as \textit{deflection} between the ``noise only'' and
  ``signal+noise'' hypothesis. The deflection is homogeneous to the signal-to-noise ratio and it
  uses the quantities estimated at the first step.
  
  We apply this original method to the problem of the detection of supernovae core collapses.  We
  use the catalog of waveforms provided recently by Dimmelmeier et al. \cite{dimmelmeier02:_relat2}
  to train our algorithm. We expect such detector to have better performances on this particular
  problem provided that the reference signals are reliable.}


\section{Motivations}
A number of large scale gravitational wave interferometric detectors such as LIGO, TAMA, GEO600 and
VIRGO \cite{gwdetectors} are taking scientific data or will reach this goal soon. The objective of
this paper is to contribute to the arsenal of detection algorithms able to locate the weak and short
signature of a gravitational wave of astrophysical origin in the long data stream produced by the
detector.

In the list of candidates having a good chance for the first detection, there are sources for which
we can only make a rough guess or simulate the highly non-linear Physics which is involved. This
causes the expected gravitational waveforms to be poorly modeled. Most of these are collapses of
very massive astrophysical objects in the final stage, and this yields the resulting gravitational
wave to be a burst.

In such cases, computer simulations may give good indications of what can be the waveform for some
choices of the parameter values describing the physical phenomenon. However, it is generally not
possible to have a tight sampling of the parameter space i.e., to scan a large range of physical
configurations. We only have at our disposal a catalog of waveforms, whose members are selected
representatives of the large set of possibilities.

Two examples of such sources are supernovae core collapse and binary black hole merger. Despite some
recent progresses, work is still needed in the latter case to produce reliable waveforms in a
realistic set up (including spins for instance). Concerning the former case, hydrodynamic
simulations of relativistic supernovae have been recently computed
\cite{dimmelmeier02:_relat2,dimmelmeier02:_relat1} and the expected waveforms for different
parameter configurations were made available.

In this paper, we propose a method for designing systematically a decision statistic for the
detection of gravitational transients by extracting the necessary information from a catalog of test
waveforms emitted from a targeted source. We use the supernovae waveforms as one possible
application.  Although not considered in this paper, the presented approach may also apply to other
problems encountered when analyzing the output of a gravitational wave interferometer, such as the
classification and the characterization of the noisy transients. (Because they worsen the detector
sensitivity, such interferences deserve the development of algorithms to determine their actual
origin.)  In this context, the initial database could be a collection of characteristic individuals
extracted ``by hand'' or with some other simple algorithm. In any case, we refer the (gravitational
wave or noise) transient(s) we want to detect to as \textit{signal}.

Some attention has been paid to the choice of an adequate vector formalism to treat the problem and
make the implementation on computers easier. The resulting notations are defined in Sect.
\ref{notation}. This section also includes the formulation, within the chosen framework, of
classical results such as the Plancherel formula which will be of use further.

In Sect. \ref{probstatement}, we describe the detection problem we consider with the accompanying
hypothesis. We assume the signal to be random and of unknown probability density function (PDF).
This assumption translates explicitly the lack of knowledge about the signal. The only piece of
information at our disposal is its first and second order statistical moments (i.e., its mean and
covariance). In the situation of interest here, these two quantities are not known, but they can be
estimated with sufficient accuracy from available sample sets.

We consider that the noise is Gaussian and stationary and that we know its correlation function (or
equivalently its power spectral density). Analogously to the signal, an extension to the case where
there is no reasonable noise model, is possible through the use of estimates done with ``noise only''
data streams.

Since we do not have the signal PDF, it is impossible to write the exact form of the likelihood
ratio, and thus to obtain the optimal statistic. However, a satisfactory solution can be obtained
by first imposing the mathematical structure of the statistic and second look in the selected set
of functions for the best element by maximizing a contrast criterion.

The difficulty then lies in making the choice of a sufficiently general class of statistics and a
sensible criterion for the considered problem. In Sect. \ref{quadraticdetector}, it is explained why
the family of \textit{quadratic detectors} (i.e., the detection statistic is a quadratic form of the
observed data) and a measurement of the \textit{deflection} (a quantity homogeneous to the signal to
noise ratio) are good candidates. Coming back to our initial detection problem, we individuate in
the selected class the statistic which performs best according to the chosen criterion and show that
it can be expressed easily with the signal and noise covariance.  Our proof was inspired by the work
presented in \cite{matz96:_time} and \cite{picinbono88:_optim} which we adapt to the case of
interest (finite vector spaces and non-centered signals). We finish Sect. \ref{quadraticdetector}
showing that the proposed approach is not unfamiliar since it can be related to the well-known
method of matched filtering.

In Sect. \ref{application}, we put the quadratic detector of best deflection into practice with the
problem of detecting gravitational wave transients from supernovae core collapse. In this specific
case, we show how the signal covariance matrix can be estimated from the catalog of simulated
waveforms. This yields a simplification of the detector and an efficient implementation for the
online detection. We give also some results about the determination of the decision threshold
required to get a chosen false alarm rate.

The vector formalism introduced here is a general framework in which all quadratic detectors can be
easily related and compared. In Sect. \ref{other}, we do this comparative study between the solution
with optimal deflection and other detection techniques \cite{anderson01:_excess,pradier01:_effic}
proposed in the literature that are also belonging to the quadratic class.

\section{Notations and basic algebra}
\label{notation}
We will denote scalar quantities and scalar-valued functions with plain italics, e.g., $x$; vectors
by boldface letters, e.g. $\vec{x}$; and matrices by boldface capitals, e.g., $\op{X}$.  We will
represent the components of vectors and matrices with superscripts within brackets, e.g.
$\op{X}^{(m,n)}$ designates the element located in $m$th row and $n$th column of the matrix
$\op{X}$.  Finally, $\vec{x}^{t}$ denotes the transpose of the vector $\vec{x}$.  The symbol
$\equiv$ will be used in the following to define our variables and therefore stands for ``equal by
definition.''
  
We use curve brackets for denoting continuous (random and/or deterministic) time (or frequency)
series (e.g. $x(t)$), whereas squared brackets are employed for discrete time (sampled) processes
(e.g., $x[k]$). The samples of a discrete time signal are collected in a single column vector of
$\R^N$, e.g.
\begin{equation}
\vec{x}\equiv(x[k]=x(t_s k), k=0\ldots N-1)^t
\end{equation}
where $t_s \equiv 1/f_s$ is the sampling period and $f_s$ the sampling rate.

We define the Fourier transform of $x[k]$ by~:
\begin{equation}
X(f)\equiv t_s\sum_{k=-\infty}^{+\infty}x[k]e^{-2i\pi kf/f_s}.
\end{equation}

As a general rule, Fourier transforms are denoted with the same (capital) letter used for
its associated time sequence. The function $X(f)$ is $f_s$--periodic (i.e., $X(f)=X(f+f_s)$ for all
$f$) and its inverse may be calculated with the following inversion equation~:
\begin{equation}
x[k]= \int_{-f_s/2}^{f_s/2}X(f)e^{2i\pi kf/f_s}\,df.
\end{equation}

We recall that the Plancherel formula relates scalar products expressed in the time and frequency
domains~:
\begin{equation}
t_s\sum_{j=-\infty}^{+\infty}x[j]y[j]=\int_{-f_s/2}^{+f_s/2}X(f)\conj{Y(f)}\,df.
\end{equation}

This equation may also be expressed using vector scalar product, provided the assumption that one of
the two signals $x[k]$ or $y[k]$ has a finite support (denoted with $\supp\{\cdot\}$), as specified
in the following lemma~:
\begin{lemma}
\label{plancherel}
Assuming that  $\supp\{y[\cdot]\} \subset \{0 \ldots N-1\}$, the Plancherel formula writes~:
\begin{equation}
t_s\vec{x}^t\vec{y}=\int_{-f_s/2}^{+f_s/2}X(f)\conj{Y(f)}\,df
\end{equation}
\end{lemma}

Continuing along the same idea, the convolution of two signals $x[k]$ and $y[k]$ defined by~:
\begin{equation}
\label{conv_def}
z[k]=t_s \sum_{j=-\infty}^{+\infty}x[k-j]y[j]
\end{equation}
or equivalently with $Z(f)=X(f)Y(f)$, may also be rewritten using vectors under some support
constraints as in the next lemma.
\begin{lemma}
\label{convolution}
Let $N_y<(N-1)/2$ be a positive integer and suppose that $\supp\{y[\cdot]\}=\{-N_y \ldots N_y\}$, the collection
of samples $\vec{z}=(z[N_y] \,\ldots\, z[N-1-N_y])^t$ where $z[k]$ is the convolution of $x[k]$ and $y[k]$ as
defined in eq. (\ref{conv_def}), can be expressed as~:
\begin{equation}
\vec{z}=t_s \op{Y} \vec{x}
\end{equation}
where $\op{Y} \in \R^{N-2N_y \times N}$ is a matrix of the form
\begin{equation}
\op{Y}=\begin{pmatrix}
y[N_y]&\dots&y[-N_y]&0&\hdotsfor{2}&0\\
0&y[N_y]&\dots&y[-N_y]&0&\dots&0\\
\hdotsfor{7}\\
0&\hdotsfor{2}&0&y[N_y]&\dots&y[-N_y]
\end{pmatrix}.
\end{equation}
\end{lemma}

Proofs of both Lemma \ref{plancherel} and \ref{convolution} are simple and left to the reader.

\section{Problem statement}
\label{probstatement}
The question we consider here is to detect a (possibly non-stationary) random signal in a stationary
Gaussian noise (the signal and noise covariance function being known or could be estimated with
accuracy). Using the notations defined in the previous section, the problem is to distinguish between
two statistical hypothesis $(H_0)$ and $(H_1)$~:
\begin{subequations}
\label{problem}
\begin{align}
(H_0)\,:\quad \vec{x}&=\vec{n}\\
(H_1)\,:\quad \vec{x}&=\vec{s}+\vec{n} 
\end{align}
\end{subequations}
with the following assumptions
\begin{enumerate}
\item the signal $\vec{s}$ is a random vector of mean $\vec{s}_m\equiv\E[\vec{s}]$ (where $\E[\cdot]$ denotes
  the expectation operator) and correlation matrix $\op{R}_s\equiv\E[(\vec{s}-\vec{s}_m)(\vec{s}-\vec{s}_m)^t]$,
\item the noise $\vec{n}$ is a zero-mean, stationary Gaussian vector with correlation matrix
  $\op{R}_n$. Note that, since $\vec{n}$ is stationary, $\op{R}_n$ is a Toeplitz symmetric matrix
  (the terms of $\op{R}_n$ are given by the autocorrelation function
  $\op{R}_n^{(j,j+k)}\equiv\E[n[j]n[j+k]]$),
\item the signal and the noise are decorrelated, meaning that $\E[\vec{n}^t(\vec{s}-\vec{s}_m)]=0$,
\end{enumerate}

This set of assumptions correspond to several different practical situations. A first one is when
the signal or noise models are good enough to get reliable close form expressions of $\vec{s}_m$,
$\op{R}_s$ and $\op{R}_n$. The second situation is when a sufficiently large set of ``signal only''
and/or ``noise only'' realizations is available and can be used to obtain a good estimate of the
first and second order moments of the signal and the noise. Note that, except for its first and
second order moments, we made no hypothesis about the PDF of $\vec{s}$.

\section{Quadratic detectors}
\label{quadraticdetector}
Deciding $(H_1)$ or $(H_0)$ is classically done by finding a partition function $\Lambda(\cdot)$
dividing the observation space (here, $\R^N$) into two disjoint subsets~:
\begin{subequations}
\begin{align}
\Lambda(\vec{x}) &\geq \eta && \text{decide $(H_1)$}\\
\Lambda(\vec{x}) &< \eta && \text{decide $(H_0)$}
\end{align}
\end{subequations}
where the detection threshold $\eta$ defines the border between the two decision area. Its value is
given by fixing to a reasonable value the probability of deciding upon hypothesis $(H_1)$ although
no signal $\vec{s}$ is present, which we refer to ``false alarm probability''.  The function
$\Lambda(\cdot)$ is referred to as \textit{detection statistic} or simply \textit{detector}.

\subsection{Intuitive background}
\label{intuitive}
It is intuitive to search for some unknown signal by looking for abnormal excess of power in one or
several frequency bands of the observed signal spectrum. To implement this idea, we define the power
of a signal $x$ using a $l^2$ measure~:
\begin{equation}
E_x \equiv 1/N \sum_{k=0}^{N-1}x[k]^2=\vec{x}^t\vec{x}/N,
\end{equation}
and a bank of filters which select adequately the signal in the frequency bands of interest. Let
$\{g_m[k], k=0 \ldots\, N-1 \text{ and } m=0 \ldots\,M-1\}$ be the impulse responses (assumed to be of finite
support) of the chosen bandpass filters. We get the signal $y_m[k]$ at the output of each filter by
convolving the observed signal $x[k]$ to the corresponding impulse response. With the constraint
that $\supp\{g_m\} \subset \{-N_g \ldots\, N_g\}$ for all $m$ where $N_g<(N-1)/2$, we can apply the
Lemma \ref{convolution} and express the output signal in a vector form as~:
\begin{equation}
\vec{y}_m= t_s \op{G}_m \vec{x},
\end{equation}
where $\vec{y}_m$ and $\op{G}_m$ are as defined in Lemma \ref{convolution}.

Hence, we can write down the detection statistic corresponding to the basic idea mentioned above by
summing up the power in all $M$ bands which yields~:
\begin{equation}
\label{intuitive_detect}
\Lambda_{intuitive}(\vec{x})\equiv \sum_{m=0}^{M-1} E_{y_m}^2=\vec{x}^t\left(\frac{t^2_s}{N-2N_g}\sum_{m=0}^{M-1}\op{G}_m^t\op{G}_m\right)\vec{x}.
\end{equation}

We conclude that the heuristic principle we chose, is practically implemented with a detection
statistic which is a quadratic form of the data. Extending this result to cases where the kernel of
the form is an arbitrary symmetric matrix, this leads us to define the following family of
detectors~:
\begin{defin}
\label{defquaddetect}
Let $\op{A} \in \R^{N \times N}$ be a symmetric real matrix, the function
\begin{equation}
\label{quaddetect}
\Lambda_{\op{A}}(\vec{x})=\vec{x}^t\op{A}\vec{x},
\end{equation}
defines a \textit{quadratic detector} of kernel $\op{A}$.
\end{defin}

Quadratic detectors will be a central ingredient in this paper. It is worth noting that they have
been extensively used in many applications (see e.g. \cite{wang00:_perfor_study} for a list of
examples). In the case of gravitational wave detection, the specific area of interest here, several
works \cite{anderson01:_excess,vicere02:_optim,pradier01:_effic,arnaud99:_detec,mohanty00:_robus}
are based on such detector structure.

Beyond qualitative arguments, the following theoretical result is an important motivation for the
use of quadratic detectors \cite{van68:_detec}~: with the additional assumption of
a zero-mean Gaussian signal (i.e., both signal and noise are Gaussian), the optimal solution in the
Neymann-Pearson sense of the problem (\ref{problem}) belongs to the family defined in Def.
\ref{defquaddetect}. Although the problem considered here excludes the signal to be Gaussian, we
expect quadratic detectors to retain their good performances, when the signal PDF is close to
Gaussian.

\subsection{Optimal quadratic detectors}
\label{opt_quad_det}
For our problem (\ref{problem}), we don't have a complete knowledge of the statistics of the input
signal (the PDF of $\vec{s}$ is unknown). In consequence, we cannot write out the likelihood ratio which
gives the best (in the Neymann-Pearson sense) detector among all possibilities.

We propose to overcome this difficulty by first, reducing the class of possible solutions to the
family of the quadratic detectors defined in Def. \ref{defquaddetect} and second, extracting from this
smaller set the statistic which will best perform for our problem. More precisely, our objective is
to get the quadratic detector (i.e., get the kernel matrix $\op{A}$ which identifies this quadratic
detector in the whole family) which maximizes the following contrast criterion based on the first
and second order moments~:
\begin{equation}
\label{deflection}
d^2(\Lambda_{\op{A}})=\frac{(\E[\Lambda_{\op{A}}(\vec{x})|H_1]-\E[\Lambda_{\op{A}}(\vec{x})|H_0])^2}
{\var\{\Lambda_{\op{A}}(\vec{x})|H_0\}}.
\end{equation}

This criterion is generally referred to as \textit{signal-to-noise ratio} (statistics) or
\textit{deflection} (signal processing). The terminology ``signal-to-noise ratio'' is generally
associated in most of the literature about the gravitational wave data analysis to the quantity in
eq. (\ref{deflection}) where $\Lambda(\cdot)$ is set to the matched filter statistic. In
consequence, we adopt the term ``deflection'' to avoid confusion. The deflection may be viewed as a
contrast measurement between the two statistical hypothesis $H_0$ and $H_1$ in the sense it measures
the distance between the centers of the PDF of the statistic in the two hypothesis relatively to the
PDF width in the null hypothesis $H_0$.

In the context of the problem (\ref{problem}), we apply now this approach for selecting the best statistic among
all quadratic detectors.

\begin{lemma}
\label{quaddeflection}
  In the situation described in eq. (\ref{problem}), the deflection of a quadratic detector as
  defined in Def. \ref{defquaddetect} is
\begin{equation}
d^2(\Lambda_{\op{A}})=\frac{\trace^2\{\op{A}\op{C}_s\}}{2\trace\{(\op{A}\op{R}_n)^2\}},
\end{equation}
where $\op{C}_s\equiv\E[\vec{s}^t\vec{s}]$ defines the (non-central) covariance matrix and
$\trace\{\cdot\}$ is the trace \footnote{Let $\op{A} \in \R^{N \times N}$, the
  operator $\trace\{\op{A}\}\equiv\sum_{n=0}^{N}\op{A}^{(n,n)}$ defines the trace of $\op{A}$.}
operator.
\end{lemma}
\begin{proof}
  The proof of this lemma may be found in other articles \cite{matz96:_time,hlawatsch98} for
  zero mean signals and infinite vector spaces. Here, we give the proof for non-central signals
  (i.e, $\vec{s}_m\neq\vec{0}$) and in the case of discrete signals forming vectors of finite size.
  
  We compute the first two statistical moments of $\Lambda_{\op{A}}(\vec{x})$ under hypothesis
  $H_{0}$ and $H_{1}$. Starting conditionally to $H_0$, we have
  \begin{equation}
    \E[\Lambda_{\op{A}}(\vec{x})|H_{0}]=\E[\vec{n}^t\op{A}\vec{n}].
  \end{equation}
  
  Using the identity $\vec{x}^t\vec{x}=\trace\{\vec{x}\vec{x}^t\}$, this can be reduced in
  \begin{equation}
    \E[\Lambda_{\op{A}}(\vec{x})|H_{0}]=\trace\{\op{A}\E[\vec{n}\vec{n}^t]\}=\trace\{\op{A}\op{R}_n\}.
  \end{equation}
  
  Under the ``signal+noise'' $H_1$ hypothesis, we expand the quadratic form
  \begin{equation}
    \E[\Lambda_{\op{A}}(\vec{x})|H_{1}]=\E[\vec{s}^t\op{A}\vec{s}+\vec{s}^t\op{A}\vec{n}+\vec{n}^t\op{A}\vec{s}+\vec{n}^t\op{A}\vec{n}]
  \end{equation}
  
  Then, with the identity mentioned previously, we can simplify the auto-terms in the expansion
  $\E[\vec{s}^t\op{A}\vec{s}]=\trace\{\op{A}\op{C}_s\}$ with $\op{C}_s\equiv\E[\vec{s}^t\vec{s}]$
  while the cross terms vanish~: $\E[\vec{n}^t\vec{s}]=\E[\vec{n}^t]\vec{s}_m=0$ and
  $\E[\vec{n}^t\op{A}\vec{s}]=\E[\vec{s}^t\op{A}\vec{n}]=0$, which yields
  \begin{equation}
    \E[\Lambda_{\op{A}}(\vec{x})|H_{1}]=\trace\{\op{A}(\op{C}_s+\op{R}_n)\}.
  \end{equation}
    
  For the general expressions of higher order moments of Gaussian quadratic forms in
  \cite{johnson70:_distr_statis}, we get the variance under $H_0$~:
  \begin{equation}
    \var(\Lambda_{\op{A}}(\vec{x})|H_{0})=\var(\vec{n}^t\op{A}\vec{n})=2\,\trace\{(\op{A}\op{R}_n)^2\}.
  \end{equation}
  
  The combination of all these ingredients leads to the result.
\end{proof}

In order to find the best detector in the quadratic family, we need now to find which kernel matrix
maximizes the deflection. This is stated by the following proposition.

\begin{prop}
  There exists a unique symmetric matrix $\op{H}$ such that $d^2(\Lambda_{\op{H}})$ obtained in
  Lemma \ref{quaddeflection} is maximum, and this matrix is
\begin{equation} 
\label{optkernel}
\op{H}= \argmax_{\op{A}}\{d^2(\Lambda_{\op{A}})\} = \op{R}_n^{-1}\op{C}_s\op{R}_n^{-1},
\end{equation}
\end{prop}
\begin{proof}
  
  We first note that the noise autocorrelation $\op{R}_n$ is symmetric positive definite matrix.
  Therefore, there exists a triangular matrix $\op{T}_n$ with positive diagonal such that
  $\op{R}_n=\op{T}_n\op{T}_n^t$. This factorization method is referred to as Cholesky factorization
  \cite{golub89:_matrix}.
  
  It is useful to introduce the two following matrices $\op{G}\equiv\op{T}_n\op{A}\op{T}_n^t$ and
  $\op{C}\equiv\left(\op{T}^{-1}_n\right)^t\op{C}_s\op{T}_n^{-1}$, which we make appear in the expression
  of the deflection we got in Lemma \ref{quaddeflection}, then reducing to~:
\begin{equation}
d^2(\Lambda_{\op{A}})=\frac{\trace^2\{\op{G}\op{C}\}}{2\,\trace\{\op{G}^2\}}.
\end{equation}

Let ${\cal S}_N(\R)$ be the vector space of real symmetric matrices of $\R^{N \times N}$.  It is
easily shown that $\lag\op{A},\op{B}\rag_{{\cal S}_N(\R)}\equiv\trace\{\op{A}\op{B}\}$ defines a
scalar product on ${\cal S}_N(\R)$. Since the matrices $\op{G}$ and $\op{C}$ belong to ${\cal
  S}_N(\R)$, we can rewrite the deflection as a ratio of scalar products, also referred to as
Rayleigh quotient \cite{golub89:_matrix}, namely~:
\begin{equation}
d^2(\Lambda_{\op{A}})=\frac{\lag\op{G},\op{C}\rag^2_{{\cal S}_N(\R)}}{2\lag\op{G},\op{G}\rag_{{\cal S}_N(\R)}}.
\end{equation}

Using the Cauchy-Schwarz inequality, we conclude that $d^2(\Lambda_{\op{A}})$ is maximum if and only
if $\op{G}\propto\op{C}$. Setting $\op{G}=\op{C}$ without loss of generality and replacing $\op{G}$
and $\op{C}$ by their definition directly yields eq.  (\ref{optkernel}).
\end{proof}

We now have the final expression of the quadratic detector reaching the deflection optimum, namely
\begin{equation}
\label{optdetect}
\Lambda_{\op{H}}(\vec{x})=\vec{x}^t\op{H}\vec{x} \quad \text{given  $\op{H}=\op{R}_n^{-1}\op{C}_s\op{R}_n^{-1}$}.
\end{equation}

Before looking how the approach proposed here may be practically implemented, we first give some
interpretations of the detection statistic we obtained.

\subsection{Interpretation and relation to matched filtering}
\label{interpretation}
With a direct calculation from its definition, we can separate $\op{C}_s\equiv\E[\vec{s}^t\vec{s}]$
into two terms $\op{C}_s=\vec{s}_m\vec{s}_m^t+\op{R}_s$. The mean $\vec{s}_m$ can be viewed as a
trend of the signal which happens systematically.  In this sense, we refer the first term to as
``deterministic''. The correlation matrix is related to the typical amplitude of the
random fluctuations superimposed to the mean. For this reason, we refer the second term to as
``random''.

Injecting this expansion in (\ref{optdetect}), we obtain a similar separation of $\Lambda_{\op{H}}(\vec{x})$
\begin{equation}
\Lambda_{\op{H}}(\vec{x})=\Lambda^{det}_{\op{H}}(\vec{x})+\Lambda^{rand}_{\op{H}}(\vec{x}),
\end{equation}
where $\Lambda^{det}_{\op{H}}(\vec{x})=(\vec{s}_m^t\op{R}_n^{-1}\vec{x})^2$ is related to the
deterministic part of the signal model and
$\Lambda^{rand}_{\op{H}}(\vec{x})=\vec{x}^t\op{R}_n^{-1}\op{R}_s\op{R}_n^{-1}\vec{x}$ to its random
part.

The two contributions of the detection statistic are worth to be investigated further. An
interesting interpretation and a link to matched filtering \cite{van68:_detec} results from the
reformulation in the frequency domain of $\vec{y}^t\op{R}_n^{-1}\vec{x}$ where $\vec{x}$ and
$\vec{y} \in \R^{N}$. This is the objective of the Proposition whose proof is detailed in the next
section.

\subsubsection{Towards matched filtering}
As the preamble, we define formally the whitening operation (i.e., filtering the signal with the
inverse of the square root of the noise power spectral density) in the vector formalism used here.  Let
$\Gamma_n(f)\equiv\E[|N(f)|^2]$ be the power spectral density of the noise $n[k]$
and $\whiten{x}[k]$ be the result of the whitening of a given signal $x[k]$, we define
\begin{equation}
\label{whiten}
\whiten{X}(f)\equiv\frac{X(f)}{\sqrt{\Gamma_n(f)}}.
\end{equation}

Clearly, $\whiten{x}[k]$ is the result of the convolution of the signal $x[k]$ with the whitening
filter of impulse response $w[k]$,
\begin{equation}
\label{whiten_filt}
w[k]=\int_{-f_s/2}^{+f_s/2}{\frac{1}{\sqrt{\Gamma_n(f)}}e^{2\pi ikf/f_s}\,df}.
\end{equation}

Using the Lemma \ref{convolution}, we deduce that the whitening filter defined in (\ref{whiten}) can
be written in the vector formalism with
\begin{equation}
\label{whiten_vect}
\whiten{\vec{x}}=t_s \op{W}\vec{x},
\end{equation}
where $\whiten{\vec{x}}$ and $\op{W}$ are as given in Lemma \ref{convolution}.

This expression of the whitening operation is needed in the proof of the following Proposition, in
which we get the vector form of the matched filtering statistic provided some care for the cancellation
of finite size effects.

\begin{prop}
  \label{whitenscalprod} 
  Let $N_w<(N-1)/2$ be the half-size of $\supp\{w\}=\{-N_w\,\ldots\,N_w\}$ i.e., the support of the
  impulse response of the whitening filter defined in eq. (\ref{whiten_filt}), and let $y[k]$, a
  signal whose whiten version has a finite size support,
  $\supp\{\whiten{y}\}\subset\{0\,\ldots\,N-1\}$, then
  \begin{equation}
    \vec{y}^t\op{R}_n^{-1}\vec{x}=\int_{-f_s/2}^{+f_s/2}\frac{X(f)\conj{Y(f)}}{\Gamma_n(f)}\,df.
  \end{equation}
\end{prop}
\begin{proof}
  We first compute $\E[\whiten{\vec{n}}\whiten{\vec{n}}^t]$ and get a first expression using eq.
  (\ref{whiten_vect})~:
\begin{equation}
\label{covwithw}
\E[\whiten{\vec{n}}\whiten{\vec{n}}^t]=t_s^2\op{W}\op{R}_n\op{W}^t.
\end{equation}

A second expression is obtained by writing each terms of the considered matrix in the Fourier
domain. The component located at the $j$th row and $k$th column may be expressed as
\begin{equation}
\E[\whiten{n}[j]\whiten{n}[k]]=\iint_{-f_s/2}^{+f_s/2}{\frac{\E[N(f)\conj{N(f')}]}{\sqrt{\Gamma_n(f)\Gamma_n(f')}}e^{2\pi i (jf-kf')/f_s}\,dfdf'}.
\end{equation}

The integrand can be work out applying the Wiener-Khinchine theorem \cite{van68:_detec}~:
\begin{equation}
\label{wk}
\E[N(f)\conj{N(f')}]=\delta(f-f')\Gamma_n(f)
\end{equation}
where  $\delta(f)=t_s\sum_{k=-\infty}^{+\infty}e^{-2\pi i kf/f_s}$. 

The function $\delta$ acts on the elements of the set ${\cal P}(f_s)$ (containing the periodic
functions of period $f_s$) the same way the Dirac operator acts on functions of $L^2(\R)$. Let
$\Phi$ be a test function of ${\cal P}(f_s)$ and $\phi$ its inverse Fourier transform, we have
\begin{equation}
\label{dirac}
\int_{-f_s/2}^{+f_s/2}\Phi(f)\delta(f)\,df=t_s\sum_{k=-\infty}^{+\infty}\phi[k]=\Phi(0).
\end{equation}
 
Note also that $\delta$ is $f_s$--periodic i.e., $\delta(f+f_s)=\delta(f)$, for all $f$.  The proofs
of these two properties of $\delta$ are left to the reader. 

With (\ref{wk}), we have
\begin{equation}
\E[\whiten{n}[j]\whiten{n}[k]]=\int_{-f_s/2}^{+f_s/2}{\frac{e^{2\pi i jf/f_s}}{\sqrt{\Gamma_n(f)}}\left(\int_{-f_s/2}^{+f_s/2}\sqrt{\Gamma_n(f')}\delta(f-f')e^{-2\pi i kf'/f_s}\,df'\right)\,df},
\end{equation}
which simplifies with the change of variables $u=f-f'$ and using the periodicity of the functions
$\delta$ and $\Gamma_n$,
\begin{equation}
\E[\whiten{n}[j]\whiten{n}[k]]=\int_{-f_s/2}^{+f_s/2}{\frac{e^{2\pi i  (j-k)f/f_s}}{\sqrt{\Gamma_n(f)}}\left(\int_{-f_s/2}^{+f_s/2}\sqrt{\Gamma_n(f-u)}\delta(u)e^{2\pi i k u/f_s}\,du\right)\,df},
\end{equation}
leading finally using (\ref{dirac}) to
\begin{equation}
\E[\whiten{n}[j]\whiten{n}[k]]=\int_{-f_s/2}^{+f_s/2}{e^{2\pi i (j-k)f/f_s}\,df},
\end{equation}
or equivalently to
\begin{equation}
\E[\whiten{n}[j]\whiten{n}[k]]=f_s \delta_{jk}
\end{equation}
where $\delta_{jk}$ is the Kronecker symbol (by definition, $\delta_{jk}=1$ if $j=k$, 0 otherwise).

We conclude that 
\begin{equation}
\label{directcov}
\E[\whiten{\vec{n}}\whiten{\vec{n}}^t]=f_s \op{I}.
\end{equation}

Assuming that $\op{R}_n$ is invertible and combining both eqs. (\ref{covwithw}) and
(\ref{directcov}), we get the relation between the whitening operator and the noise correlation
matrix, namely
\begin{equation}
\label{doublewhiten}
\op{R}^{-1}_n=t_s^3\,\op{W}^t\op{W}.
\end{equation}

With eq. (\ref{whiten_vect}) and the Plancherel formula in eq. (\ref{plancherel})
\begin{equation}
\vec{y}^t\op{R}_n^{-1}\vec{x}=t_s\whiten{\vec{y}}^t\whiten{\vec{x}}=\int_{-f_s/2}^{+f_s/2}{\whiten{X}(f)\conj{\whiten{Y}(f)}\,df},
\end{equation}
provided that $\whiten{y}[k]$ has support in $\{0\,\ldots\,N-1\}$. Replacing $\whiten{Y}(f)$ and
$\whiten{X}(f)$ by their definitions completes the proof.
\end{proof}

\subsubsection{Deterministic and random components}
If we choose $N$ large enough so that no finite size effect appears (i.e., the supports of all
required signals respect the condition of Prop. \ref{whitenscalprod}), we can rewrite the
deterministic term of the detection statistic as
\begin{equation}
\label{det}
\Lambda^{det}_{\op{H}}(\vec{x})=\left(\int_{-f_s/2}^{+f_s/2}\frac{X(f)\conj{S_m(f)}}{\Gamma_n(f)}\,df\right)^2.
\end{equation}

From the following eigen expansion of the signal correlation matrix
\begin{equation}
\op{R}_s=\sum_{k=0}^{N-1}\sigma_k \vec{v}_k \vec{v}^t_k,
\end{equation}
the random component may be expressed as~:
\begin{equation}
\Lambda^{rand}_{\op{H}}(\vec{x})=\sum_{k=0}^{N-1}\sigma_k (\vec{v}_k^t\op{R}_n^{-1}\vec{x})^2.
\end{equation}

Similarly to the deterministic component, assuming again that $N$ is sufficiently large, it follows that~: 
\begin{equation}
\label{rand}
\Lambda^{rand}_{\op{H}}(\vec{x})=\sum_{k=0}^{N-1}{\sigma_k \left(\int_{-f_s/2}^{+f_s/2}\frac{X(f)\conj{V_k(f)}}{\Gamma_n(f)}\,df\right)^2}
\end{equation}

Eqs. (\ref{det}) and (\ref{rand}) show that the quadratic detector with optimal deflection
$\Lambda_{\op{H}}(\vec{x})$ is closely related to the well-known technique of matched filtering
\cite{van68:_detec}. The complete statistic can be equivalently implemented as a bank of $N+1$
matched filters (using the templates given by $\vec{s}_m$ and $\vec{v}_k$, $k=0\ldots N-1$) whose
output energies are combined with a weighted sum.  Being a covariance matrix, $\op{R}_s$ is positive
definite. All eigenvalues $\sigma_k$ are then real and positive numbers. In consequence, the weighs
(equaled to the eigenvalues) put favor or attenuate the contribution of a corresponding term in the
summation.

\section{Application to the detection of gravitational wave transients}
\label{application}
Because the physics driving supernovae explosions is highly non-linear, the expected gravitational
radiation is difficult to obtain through analytical means. However, numerical simulations are
available \cite{dimmelmeier02:_relat2,dimmelmeier02:_relat1,zwerger97:_dynam} and a catalog of the
reference waveforms associated to typical situations is accessible on the Internet. The waveforms of
this catalog, we refer to as DFM (i.e., the initial letters of authors' names, Dimmelmeier, Font and
M\"uller) present an intrinsic diversity which motivates to look at them as if they were produced by
a single random mechanism.

Consequently, the detection problem we face is similar to the one exposed in eqs. (\ref{problem})
provided that the second order statistics of both signal $s(t)$ and noise $n(t)$ are known.
Strictly speaking, the covariance matrix $\op{C}_s$ of the signal is not available but if we assume
that the DFM gravitational waveforms are noise-free and independent realizations of the random
process $s(t)$, they can be used to get a sufficiently accurate estimate.

We can then apply the method proposed in Sect. \ref{opt_quad_det} to optimally detect $s(t)$.  From
the signal covariance estimate and a realistic noise model, we can calculate the quadratic detector
with best deflection.

\subsection{Finding the best quadratic detector}
\label{estimate_sig_covariance}
From the database publicly available on the Internet \cite{dimmelmeier02:_relat2}, we have extracted
the $N_z=25$ waveforms, which we have resampled at the constant rate of $f_s=20$ kHz. Actually three
sets of waveforms can be used~: the one drawn from a Newtonian simulation set up and the other from
a fully relativistic code. We preferred to use the latter since the result is likely to be closer to
reality. The waveforms are stored in column vectors $\{\vec{z}_k\}_{k=0 \ldots N_z-1}$ of size $N=2251$
rows. This corresponds to a maximum burst duration of about $112.5$ ms.
 
\begin{figure}
  \centerline{\includegraphics[width=\textwidth]{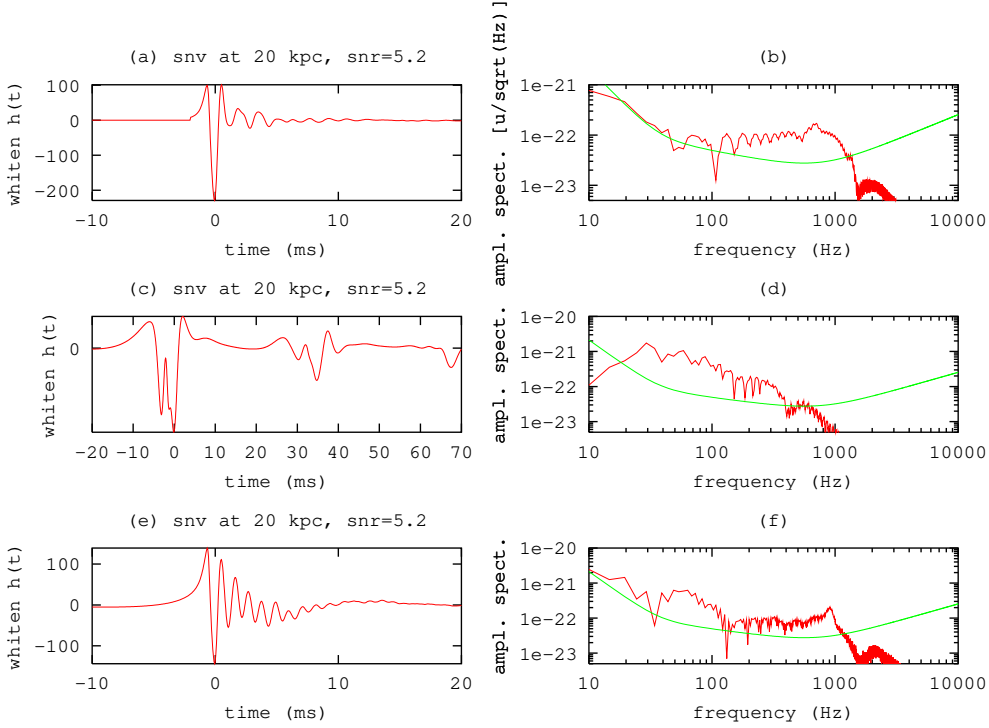}}
  \caption{\label{wvfex}\textbf{Examples of simulated gravitational transient emitted by a supernovae core collapse 
      taken from the DFM catalog}.  The DFM catalog of supernovae gravitational transients can be
    separated into three types {\protect \cite{dimmelmeier02:_relat2}} which correspond to
    different collapse scenarios. In each of these cases, we present the waveform (column on the left hand
    side) which has been filtered by the whiten filter (in eq. ({\protect \ref{whiten}})) and the
    Fourier transform of the corresponding (non-whiten) waveform (column on the right hand side)
    superimposed to the objective spectral density of Virgo noise. Each supernovae has been placed
    at a distance of $d=20$ kpc from earth which corresponds to a signal-to-noise ratio (averaged
    value obtained with all waveforms in the catalog) of about $\mathrm{SNR}=5.2$. Top row
    (\textbf{a}) and (\textbf{b})~: ``regular collapse'' (model reference: A1B3G3). Middle row
    (\textbf{c}) and (\textbf{d})~: ``multiple bounce collapse'' (model reference: A2B4G1). Bottom row
    (\textbf{e}) and (\textbf{f})~: ``rapid collapse'' (model reference: A1B3G5). The waveforms have
    clearly different shapes and characteristics (time duration and frequency bandwidth).}
\end{figure}

For each signal, we fixed the time axis origin to be located at the minimum of the largest negative
bump of the whiten signal (in most cases, this is synchronized with the core bounce
\cite{dimmelmeier02:_relat2}), precisely~: $t_0=\text{cst}=\argmin_j(\whiten{z}_k[j]), \forall k$.
Figure \ref{wvfex} presents three individuals of each type of supernovae taken from the processed DFM
catalog .

Assuming that the DFM gravitational waveforms are noise-free and independent realizations of the
random process $s(t)$, we use these waveforms to estimate the covariance matrix of $s(t)$. This is
done with the following empirical unbiased estimator~:
\begin{equation}
\label{approxcov}
\hat{\op{C}}_{s}=\frac{1}{N_z}\sum_{k=0}^{N_z-1}\vec{z}_k\vec{z}_k^t.
\end{equation}

For simplicity, we consider that the noise power spectrum is known \textit{a priori} and is given by
the expected sensitivity curve for the planned detectors. (Note that the noise correlation matrix
may also be estimated from ``noise only'' data streams.) We restrict our study to the case of the
Virgo detector using the noise model available at the following address \cite{02:_virgo}.
Extensions to other large scale interferometers are straightforward. We can get the noise
correlation matrix from the power spectrum applying the inverse Fourier transformation. From the
obtained values of $\hat{\op{C}}_{s}$ and $\op{R}_n$, we deduce the kernel of the best quadratic
detector as given in eq. (\ref{optkernel}).

This computation requires $O(N^3)$ operations to get the inverse of $\op{R}_n$ and we need roughly
the same number of operations to make the two matrix multiplications in eq. (\ref{optkernel}).  A more
computationally efficient algorithm may be used. With this aim in view, we process the whole catalog
of waveforms with the following operation $\overwh{\vec{z}}_k \equiv \op{R}_n^{-1} \vec{z}_k$.
Roughly speaking, the multiplication by $\op{R}_n^{-1}$ is equivalent to whitening the signal twice
as suggested by the factorization of $\op{R}_n^{-1}$ in eq. (\ref{doublewhiten}). The more precise
relation $\overwh{Z}_k(f)= t_s Z_k(f)/\Gamma_n(f)$ can be stated by applying the result in Prop.
\ref{whitenscalprod} with $y[j]=\delta_{jm}$ for any $m\in \{0 \ldots N-1\}$ and $x[j]=z_k[j]$. This
double whitening operation amounts to filtering the signal in the frequency bandwidth of interest
where the noise is low and removing the remaining part where the noise is large. From eq.
(\ref{optdetect}) and (\ref{approxcov}), it is easily shown that the objective kernel can be
computed directly using the modified catalog with the relation~:
\begin{equation}
\label{estimkernel}
\hat{\op{H}}=\frac{1}{N_z}\sum_{k=0}^{N_z-1}\overwh{\vec{z}}_k\overwh{\vec{z}}_k^t.
\end{equation}

Since the correlation matrix $\op{R}_n$ is Toeplitz symmetric, the computation of
$\overwh{\vec{z}}_k$ is equivalent to solving a $N_z \times N_z$ Toeplitz linear system. This can be
done efficiently with a variety of fast $O(N^2)$ algorithms. We selected and applied the Levinson
algorithm \cite{golub89:_matrix}.

The total gravitational energy radiated during the collapse varies according to the selected models.
The peak amplitudes of the waveforms of the DFM catalog have values ranging in an interval as large
as one order of magnitude. To ensure that all types of supernovae are treated equitably in the sum
of (\ref{estimkernel}), we scale all $\overwh{\vec{z}}_k$ by dividing by the expected
signal-to-noise ratio defined as~:
\begin{equation}
\mathrm{SNR}_k\equiv\left(\int_{-f_s/2}^{f_s/2}{|Z_k(f)|^2/\Gamma_n(f)\,df}\right)^{1/2}.
\end{equation}

A practical expression of this quantity can be obtained by first deducing from Prop.
\ref{whitenscalprod} that $\mathrm{SNR}_k=(\vec{z}_k^t\op{R}_n^{-1}\vec{z}_k)^{1/2}$ and using the
definition of the whitening operator in eq. (\ref{whiten_vect}) and its relation to $\op{R}_n$ in
eq. (\ref{doublewhiten}) thus leading to $\mathrm{SNR}_k=\norm{\whiten{\vec{z}}_k}_2/\sqrt{f_s}$
where $\norm{\vec{x}}^2_2\equiv\vec{x}^t\vec{x}$ defines the $l_2$ norm.

At this point, it is worth noting that eq. (\ref{estimkernel}) implements a learning scheme which
extracts systematically the necessary information from a (possibly large and heterogeneous) database
of a reference waveforms in order to find the best detector among a common used class of
possibilities.

\subsection{Approximated detector}
A close look to the detector kernel $\hat{\op{H}}$ indicates that it is degenerated (its rank is much
smaller than $N$). There are two reasons for that~: first, as a result of the linear combining of
$N_z\ll N$ rank-1 matrices (see eq. (\ref{estimkernel})), the rank of the kernel cannot exceed
$N_z$. The second reason is the fact that the waveforms of the DFM catalog have common features in
their shapes (e.g., fundamental oscillation frequency, time duration, \ldots).  This causes the
matrices $\overwh{\vec{z}}_k\overwh{\vec{z}}_k^t$ to share some linear dependency.

Precisely, it means that the kernel may be decomposed along a small number of preferred directions
of the measurement space. The most adequate basis to check this is formed by the generalized
eigen-vectors of $\hat{\op{C}}_s$ and $\op{R}_n$ as explained in the following Section. We show that
the kernel degeneracy may be used to simplify the detector and reduce its computational complexity.

\subsubsection{Truncating to principal directions}
The vector $\vec{u}$ and scalar $\gamma$ are respectively the generalized eigen-vector and value of
$\hat{\op{C}}_s$ and $\op{R}_n$ if the following equation is satisfied \cite{golub89:_matrix}~:
\begin{equation}
\label{geneigeq}
\hat{\op{C}}_s \vec{u} = \gamma \op{R}_n \vec{u}.
\end{equation}

Since $\op{R}_n$ is a definite positive matrix, it can be decomposed using the Cholesky
factorization \cite{golub89:_matrix} as the product of invertible and triangular matrices, namely
$\op{R}_n=\op{T}_n\op{T}^t_n$. Multiplying to the left both sides of (\ref{geneigeq}) by
$\op{T}^{-1}_n$, the generalized eigen problem above turns out to be equivalent to the standard one
given by~:
\begin{equation}
\opgreek{\Gamma} \vec{v} = \gamma \vec{v},
\end{equation}
provided that $\opgreek{\Gamma}=\op{T}^{-1}_n \hat{\op{C}}_s \op{T}^{-t}_n$ and
$\vec{v}=\op{T}^{t}_n\vec{u}$. Consequently, the matrix $\opgreek{\Gamma}$ may be expanded along its
eigen-directions $\{\vec{v}_k\}_{k=0 \ldots N-1}$, namely~:
\begin{equation}
\opgreek{\Gamma}=\sum_{k=0}^{N-1} \gamma_k \vec{v}_k \vec{v}^t_k.
\end{equation}

Since we have $\hat{\op{H}}=\op{T}^{-1}_n \opgreek{\Gamma} \op{T}^{-t}_n$, the previous expansion yields
the one of the detector kernel along the generalized eigen-basis defined in eq. (\ref{geneigeq})
\begin{equation}
\label{detecteigen}
\hat{\op{H}}=\sum_{k=0}^{N-1} \gamma_k \vec{u}_k \vec{u}^t_k.
\end{equation}

\begin{figure}
 \centerline{\includegraphics[width=\textwidth]{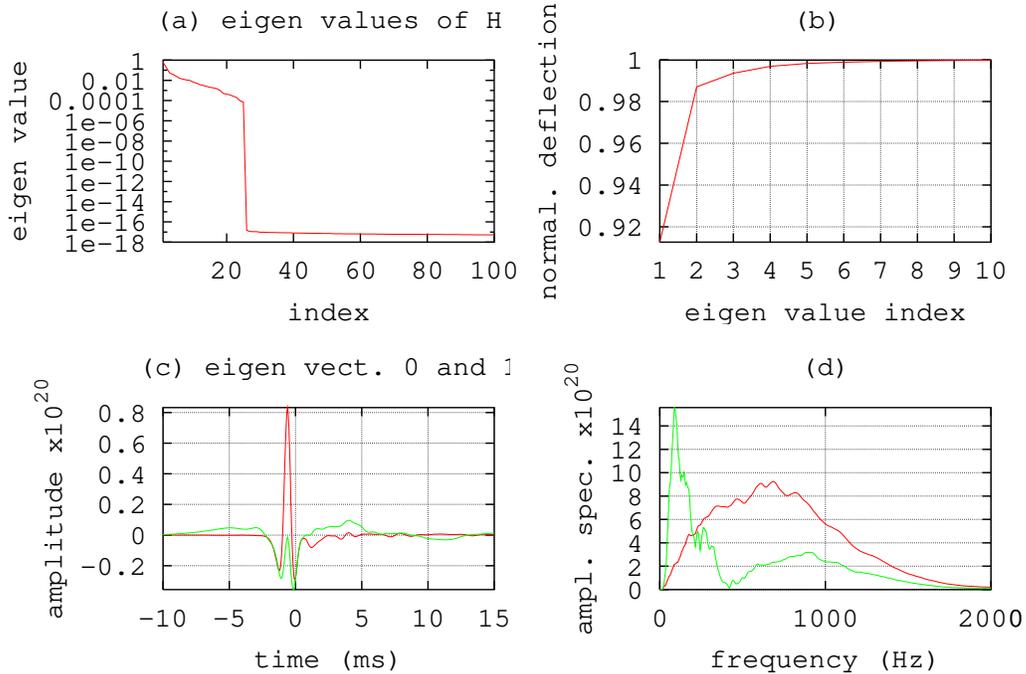}}
 \caption{\label{kereig}\textbf{Generalized eigen-values and eigen-vectors of $\hat{\op{H}}$}. With 
    these plots, we summarize the information carried by the generalized eigen-values
    $\gamma_k$ and vectors $\vec{u}_k$ defined by eqs. ({\protect \ref{geneigeq}}). The generalized
    spectrum of $\hat{\op{H}}$ is largely dominated by a few first eigenvalues (the first 100 ones
    are shown in (\textbf{a})). This degeneracy can be used to simplify the statistic by truncating
    the eigen expansion ({\protect \ref{detecteigen}}) to the first terms (see eq. ({\protect
      \ref{trunc_ker}})). The number of terms to keep is given by the amount of deflection we
    tolerate to lose due to this truncation. This is indicated in (\textbf{b}) where we see that
    keeping the $n=2$ dominating eigen-vectors is sufficient to reach $\approx 99\%$ of the optimal
    deflection. These two eigenvectors $u_0[k]$ (dark gray) and $u_1[k]$ (light gray) are presented
    in (\textbf{c}) with their respective Fourier transform in (\textbf{d}). From its shape, it
    appears that $u_0[k]$ grabs most of the peak occurring in the bounce phase of the supernovae
    (this represents about $91\%$ of the total deflection) and $u_1[k]$ the few oscillations of the
    ringdown phase (which are the $8\%$ remaining). It is worth noting that both $U_0(f)$ and
    $U_1(f)$ are non-zero in frequency bands ranging from $200$ Hz to $1$ kHz and from $50$ Hz
    to $100$ Hz.}
\end{figure}

Combining adequately a Cholesky and a Schur decomposition \cite{golub89:_matrix,02:_gnu_octav}, we
computed the solutions of eq. (\ref{geneigeq}). The eigen-values are sorted in decreasing order
$\gamma_0>\gamma_1>\ldots>\gamma_{N-1}$ and presented in Fig. \ref{kereig}.  It appears clearly that
the resulting spectrum is essentially dominated by a first few eigenvalues.

A consequence is that the sum in eq. (\ref{detecteigen}) can be fairly approximated by the summation
truncated to the first terms. Let $n<N$ be the truncation limit, we get the following kernel
\begin{equation}
\label{trunc_ker}
\tilde{\op{H}}_n\equiv \sum_{k=0}^{n-1} \gamma_k \vec{u}_k \vec{u}^t_k
\end{equation}
which we use to compute the approximated detection statistic~:
\begin{equation}
\label{trunc_det}
\Lambda_{\tilde{\op{H}}_n}(\vec{x}) = \sum_{k=0}^{n-1} \gamma_k (\hat{\vec{u}}_k^t \vec{x})^2 \approx \Lambda_{\hat{\op{H}}}(\vec{x}).
\end{equation}

The value of the truncation index $n$ is essentially related to the intrinsic complexity of the initial waveform
database. In the case of interest, $n$ is much smaller than $N$ by several order of magnitude (a
non-empirical choice of $n$ is described in the next section), and its value remains stable when $N$
increases. In consequence, the approximated statistic (\ref{trunc_det}) is computed with $O(N^2)$
floating point operations versus a total cost of $O(N^3)$ in the non approximated case.

\subsubsection{Loss in deflection due to approximation}
The truncation to a few eigen directions causes $\Lambda_{\tilde{\op{H}}_n}(\vec{x})$ to be
sub-optimal i.e., the resulting deflection is smaller than the one obtained with
$\Lambda_{\check{\op{H}}_n}(\vec{x})$.  Two interesting questions are (\textit{i}) how much
deflection do we lose? and (\textit{ii}) can we adjust $n$ so that the loss is acceptable?  To
address these questions, it is convenient to define the \textit{loss in deflection}~: $l_n \equiv
d^2(\Lambda_{\tilde{\op{H}}_n})/d^2(\Lambda_{\hat{\op{H}}})$.  This index whose values are between 0
and 1 measures the degree of ``sub-optimality'' of the truncated detector.

Replacing $\op{A}$ in the expression of the deflection obtained in Lemma \ref{quaddeflection} with
the truncated sum in eq. (\ref{trunc_ker}), and using the fact that $\{\vec{u}_k\}_{k=0 \ldots N-1}$
form a basis which diagonalizes simultaneously $\op{R}_n$ and $\hat{\op{C}}_s$ i.e., more precisely
$\vec{u}^t_k \op{R}_n \vec{u}_j = \delta_{jk}$ and $\vec{u}^t_k \hat{\op{C}}_s \vec{u}_j =
\gamma_k \delta_{jk}$ (see \cite{golub89:_matrix} for details), a straightforward calculation
leads to
\begin{equation}
d^2(\Lambda_{\tilde{\op{H}}_n})=\sum_{k=0}^{n-1}{\gamma_k^2}.
\end{equation}

This results holds also for $n=N$ yielding the maximum value of the deflection, which we denote by
$d_{max}\equiv d^2(\Lambda_{\hat{\op{H}}})=\sum_{k=0}^{N-1}\gamma_k^2$. The loss in deflection can
then be expressed as~:
\begin{equation}
l_n=\sum_{k=0}^{n-1}{\frac{\gamma_k^2}{d_{max}}}
\end{equation}
and is presented in Figure \ref{kereig} (b). We conclude that, with $n=2$ i.e., keeping the first
two leading eigen-directions, the truncated detector has a performance index of about 99\% (1\% from
optimum).  Figure \ref{kereig} (c) details the waveforms of the two leading eigenvectors.

Provided that $u_0[k]$ and $u_1[k]$ have support in $\{0 \ldots N-1\}$ (this is the case in the example
presented here, see Fig. \ref{kereig} (c)), we can apply Lemma \ref{plancherel} and get the truncated
detector (\ref{trunc_det}) expressed in the frequency domain~:
\begin{equation}
\label{trunc_det_freq}
\Lambda_{\tilde{\op{H}}}(\vec{x}) = \gamma_0 f_s^2 \left(\int_{-f_s/2}^{f_s/2} X(f)\conj{U_0(f)}\,df\right)^2+
\gamma_1 f_s^2 \left(\int_{-f_s/2}^{f_s/2} X(f)\conj{U_1(f)}\,df\right)^2.
\end{equation}

We conclude that the detection statistic is computed by first selecting the interesting frequency
contents of the spectrum of the observed data with the two (bandpass, as shown by Figure
\ref{kereig} (d)) filters $U_0(f)$ and $U_1(f)$ and then combining the energy of the filter outputs
with a weighted sum. The weight parameters can be interpreted as ``confidence coefficients'' in
finding a (supernovae) signal in the corresponding frequency bands.

In other words, the proposed method extracts systematically from a database of reference signals
the frequency bands which need to be considered in order to maximize the deflection.

\subsubsection{Detection threshold and false alarm probability}
\label{threshold_pfa}
Under noise only $(H_0)$ assumption, the detector (\ref{trunc_det}) is a finite sum of the squares
of the random variables defined by $n_k\equiv\vec{u}_k^t\vec{n}$
\begin{equation}
\Lambda_{\tilde{\op{H}}}(\vec{n}) = \sum_{k=0}^{n-1} \gamma_k n_k^2.
\end{equation}

These variables can be easily shown to be Gaussian and zero-mean. Furthermore, since
$\{\vec{u}_k\}_{k=0\ldots N-1}$ diagonalizes the noise correlation $\op{R}_n$, we have $\E[n_j
n_k]=\delta_{jk}$, from which we conclude that $\{n_k\}_{k=0\ldots n-1}$ is a sequence of
independent and identically distributed Gaussian variables of PDF ${\cal N}(0,1)$.

Let $f(\lambda)$ be the PDF of $\Lambda_{\tilde{\op{H}}}(\vec{n})$ when there is only noise.  In the
case where $n=2$ eigen-vectors are sufficient to get a good approximation of the optimal detector,
we have for $\lambda>0$ \cite{johnson70:_distr_statis}
\begin{equation}
\label{pdfH0}
f(\lambda)=\frac{1}{2\sqrt{\gamma_0 \gamma_1}}
\exp\left(-\frac{1}{4}\left(\frac{1}{\gamma_1}+\frac{1}{\gamma_0}\right)\lambda\right)
I_0\left(\frac{1}{4}\left(\frac{1}{\gamma_1}-\frac{1}{\gamma_0}\right)\lambda\right)
\end{equation}
where $I_0(\cdot)$ is the modified Bessel function of first kind
\cite{gradshteyn80:_tables_integ_series_produc} and $f(\lambda)=0$ if $\lambda\leq 0$.

\begin{figure}
  \centerline{\includegraphics[width=.7\textwidth]{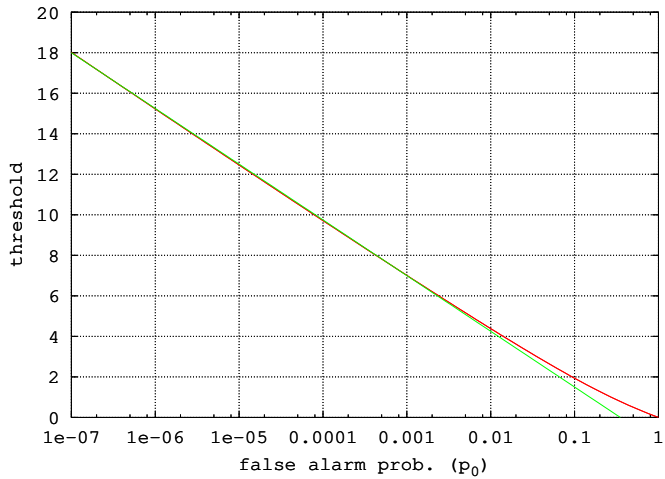}}
  \caption{\label{pfa}\textbf{Detection threshold satisfying a given requirement on the false alarm rate.} This plot is the 
    diagram (dark gray) of the function $\eta=F^{-1}(1-p_0)$ relating the detection threshold $\eta$
    to apply in order to get a fixed false alarm rate $p_0$. The function $F(\cdot)$ is the
    cumulative probability function of the statistic $\Lambda_{\tilde{\op{H}}}$ in the $H_0$ null
    hypothesis (``noise only''). It is the integral of the PDF in eq. ({\protect \ref{pdfH0}}) where
    we fixed $\gamma_0 \approx 0.627$ and $\gamma_1 \approx 0.181$. Typical values for $p_0$ are
    ranging in the interval between $10^{-7}$ to $10^{-5}$ (this roughly gives false alarm rate of a
    few per 10 mins to a few per day) which correspond to values of the threshold between $11$ and
    $18$. In this range of interest, the following fit $\eta\approx -11/4 \log_{10}(p_0)-5/4$ (light
    gray) gives a satisfactory approximation.}
\end{figure}

Integrating the PDF in eq. (\ref{pdfH0}), we obtained the cumulative probability function
$F(\lambda)\equiv\int_0^\lambda {f(\nu)\,d\nu}=\P(\Lambda_{\tilde{\op{H}}}(\vec{x})<\lambda|H_0)$.
The threshold ensuring a given false alarm probability $p_0$ is thus given by $\eta=F^{-1}(1-p_0)$.
Such function is presented in Fig. \ref{pfa} in the range of useful values of $p_0$ and using the
first two leading eigenvalues of $\hat{\op{H}}$, namely $\gamma_0 \approx 0.627$ and $\gamma_1
\approx 0.181$.

\subsubsection{Time running implementation}
Until now, we have only considered the statistical test of the presence of a supernovae transient at
a given time. The date of arrival of the gravitational wave being unknown, we must apply the
detection procedure at any given time instants. To do this, we select the $N$ data samples starting
from a given time index $m$, namely $\vec{x}_m\equiv(x[m+k], k=0\ldots N-1)^t$. We compute the
detection statistic $\Lambda_{\op{H}}(\vec{x}_m)=\vec{x}_m^t\op{H}\vec{x}_m$ for increasing and
equally spaced values of $m=0,\delta_m,\ldots$. This is similar to select the data with a time
sliding window.

Using the approximated statistic expressed as in (\ref{trunc_det_freq}) and noting that
$X_m(f)=X(f)e^{-2\pi i m f/f_s}$, we get
\begin{equation}
\Lambda_{\tilde{\op{H}}}(\vec{x}_m) = \gamma_0 f_s^2 (y_0[m])^2+\gamma_1 f_s^2 (y_1[m])^2,
\end{equation}
where $y_{(0,1)}[m]=\int_{-f_s/2}^{+f_s/2} X(f)\conj{U_{(0,1)}(f)}e^{-2\pi i mf/f_s}\,df$ are
obtained by passing the signal through a time-invariant linear filter. Assuming $U_0(f)$ and
$U_1(f)$ are stored in memory, the computation of $y_{0}[m]$ and $y_{1}[m]$ can be efficiently
computed with the FFT (and inverse) algorithm.

\section{Relation to other detection techniques}
\label{other}
We have shown in Sect. \ref{interpretation} that the quadratic detector with optimal deflection can
be related to matched filtering. In fact, many of the methods for transient detection available in
the literature, e.g.
\cite{anderson01:_excess,vicere02:_optim,pradier01:_effic,arnaud99:_detec,mohanty00:_robus} belong
to the class of quadratic detectors defined in Def. \ref{defquaddetect}. The vector formalism used here
constitutes a general framework in which all these methods can be reformulated and easily compared.
Considering that the noise model remains the same than the one we used in the previous section, we
get the shape of the kernel used by the two  contributions described in \cite{anderson01:_excess} and
\cite{pradier01:_effic} to which we limit the investigation. With this ``back-engineering''
approach, we can retrieve the \textit{a priori} assumption on the signal covariance needed for the
considered detector to have optimal deflection. We make this comparison by looking to the
generalized eigen-basis of the obtained kernel.

\subsection{Excess power statistic \cite{anderson01:_excess}} 
The basic idea is to monitor the power into one (or several) given frequency band $f_0\pm\delta f/2$
(similarly to Sect. \ref{intuitive}).  Let $X[j]\equiv\sum_{k=0}^{N-1}x[k]e^{-2\pi i jk/N}$ be the
discrete Fourier transform (DFT). The excess power statistic presented in \cite{anderson01:_excess}
reads~:
\begin{equation}
\label{eps}
\Lambda_{eps}(\vec{x})=\sum_{j=j_{-}}^{j_{+}} \frac{\abs{X[j]}^2}{\Gamma_n(j f_s/N)},
\end{equation}
where the limit indices are defined as $j_{\pm}=N(f_0\pm\delta f/2)/f_s$.

The DFT can be re-expressed as a scalar product $X[j]=\conj{\vec{f}}_j^t\vec{x}$ with the Fourier
exponentials $\vec{f}_j\equiv (e^{2\pi i jk/N}, k=0 \ldots N-1)^t$. It is straightforward to show
that the above statistic is a quadratic detector as in Def. \ref{defquaddetect} with the kernel
\begin{equation}
\op{H}_{eps}=\sum_{j=j_{-}}^{j_{+}} \frac{\vec{f}_j\conj{\vec{f}_j^t}}{\Gamma_n(j f_s/N)}.
\end{equation}

Roughly speaking, assuming the noise power spectral density is ``flat'' in the selected frequency
bandwidth, $\{\vec{f}_j\}_{j=j_{-} \ldots j_{+}}$ diagonalizes $\op{R}_n$. In this case, the kernel of
the excess power statistic has a number of $(j_{+}-j_{-}+1)=V/2$ generalized eigenvectors.  Fig.
\ref{alfeps} presents the some of these eigenvectors and their Fourier transforms.

\subsection{Linear fit filter (ALF) \cite{pradier01:_effic}}
The detection statistic $\Lambda_{alf}(\vec{x})$ is obtained from a local linear fit of
the whiten signal $\whiten{x}$.  A mean square rule yields the two parameters of the fit~:
\begin{align}
a&=\frac{\lag t\whiten{x} \rag-\lag t \rag^2}{\lag t^2 \rag-\lag t \rag^2}\\
b&=\lag \whiten{x} \rag-a\lag t \rag
\end{align}
which are orthonormalized, squared and combined to get $\Lambda_{alf}(\vec{x})$.

It turns out to be convenient to set the time origin at the center of data chunk, which we assume to
have an odd number $N$ of samples. We can do this with no lost of generality. In this set up, the
fit parameters are given by the following scalar products
$a=\vec{t}^t\whiten{\vec{x}}/\norm{\vec{t}}^2_2$ and $b=\vec{1}^t\whiten{\vec{x}}$ where we defined
$\vec{t}=(-L\,\ldots\,L)^t$, $L\equiv(N-1)/2$ being the half-size of the data chunk and
$\vec{1}=(1\,\ldots\, 1)^t$. After the orthonormalization and combining, the detection statistic
appears to be a quadratic detector as in Def. \ref{defquaddetect} of kernel
\begin{equation}
\op{H}_{alf}=t_s^2\op{W}^t \left(\frac{\vec{t}\vec{t}^t}{\norm{\vec{t}}^2_2}+\frac{\vec{1}\vec{1}^t}{\norm{\vec{1}}^2_2}\right)\op{W}
\end{equation}
where $\op{W}$ is the whitening matrix defined in eq. (\ref{whiten_vect}).

\begin{figure}
 \centerline{\includegraphics[width=\textwidth]{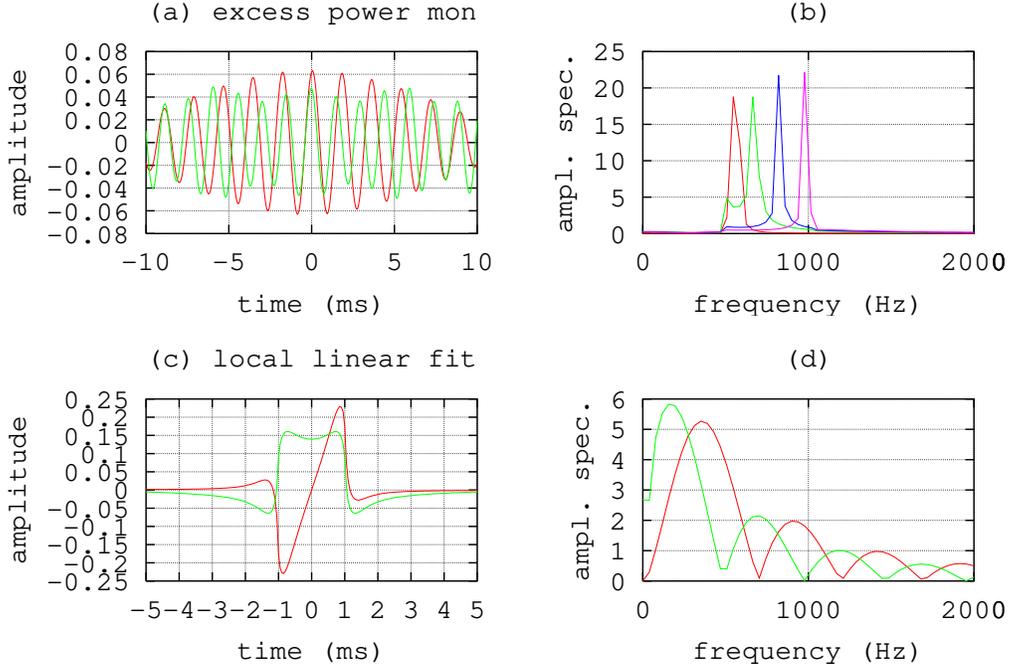}}
 \caption{\label{alfeps}\textbf{Generalized eigen-vectors of $\op{H}_{eps}$ and $\op{H}_{alf}$}.
   In this figure, we present the generalized eigenvectors (left hand side column) of the detection
   kernel used by the excess power (EPS) and the local linear fit (ALF) statistic and their
   respective Fourier transform (right hand side column). Concerning the EPS statistic (top row),
   we chosen a time window with $N=512$ samples, corresponding to a duration $\delta t \approx 25$
   ms provided a sample rate of $f_s=20$ kHz, and a frequency window of $\delta f=500$ Hz centered
   around $f_0=750$ Hz. This gives a time-frequency volume {\protect \cite{anderson01:_excess}} of
   $V \approx 2 \times 25 \mathrm{ms} \times 500 \mathrm{Hz}=25$. These parameters lead to the
   following limit indices $j_{-}=13$ and $j_{+}=26$ in the sum ({\protect \ref{eps}}). The detector
   kernel has about 14 large generalized eigenvalues which we sort in decreasing order.  The
   corresponding eigenvectors form a set of bandpass filters (width $\approx 80$ Hz) covering
   uniformly the selected frequency window. The waveforms of the 1st and the 4th eigenvectors are
   plotted in (\textbf{a}) and we show in (\textbf{b}) the spectra of the eigenvectors of rank 1, 4,
   8 and 12.  The linear fit done by the ALF statistic (bottom row) is computed using $N=40$
   samples of data (i.e., in a time window of 2 ms) which is the best time window duration found in
   {\protect \cite{pradier01:_effic}} for supernovae transients. The two eigenvectors
   $\op{W}^t\vec{t}$ (dark gray) and $\op{W}^t\vec{1}$ (light gray) are shown in (\textbf{c}) and
   their respective Fourier transform in (\textbf{d}). It appears that first filter selects
   frequencies in $350\pm 200$ Hz, and the second in $155 \pm 135$ Hz. In a sense, although the
   bandwidth are not exactly the same, this filter bank is similar to the deflection optimal
   detector.}
\end{figure}

It can be easily shown that $\op{W}^t\vec{1}$ and $\op{W}^t\vec{t}$ are the two generalized
eigenvectors of $\op{H}_{alf}$ associated to the eigenvalue 1 (this is the only non-zero
eigenvalue). They are presented in Fig. \ref{alfeps}. In this degenerated case, similar calculations
as the ones done in Sect. \ref{threshold_pfa} yields the PDF of $\Lambda_{alf}(\vec{x})$ in the
noise only case \cite{johnson70:_distr_statis}, namely~:
\begin{equation}
f_{alf}(\lambda)=\frac{1}{2}e^{-\lambda/2},
\end{equation}
from which the threshold can be obtained for a given false alarm probability. This is a
complementary contribution to the analysis made in \cite{pradier01:_effic} about the local linear
fit method.

\section{Conclusions}
Quadratic detectors (i.e., statistics which are bilinear functions of the data) can be essentially
viewed as a filtering of the data through a selection of frequency bands, the power of the filtered
data being further linearly combined. We introduced a method which systematically extracts from a
complicated and possibly large database of target signals, the important features which need to be
considered in order to design this filter bank and choose the parameters for the energy combining.
In the context of the detection of supernovae core collapses, we show that the method gives an
intuitively appealing result of a filter bank composed of two elements (the one selecting the bounce
pulse and the other, the few oscillations of the ringdown phase) whose output powers are combined in
such a way to favor the bounce (which is the most energetic part of the signal).

The scope of the approach presented here is general. The algorithm can be adapted to other problems
of the same type (for instance the detection of the final merger part of a binary black coalescence
or of non-stationary noise interferences) provided that a sufficient number of training waveforms
are available.


\end{document}